\documentclass[twocolumn,aps]{revtex4}

\usepackage{graphicx}
\usepackage{dcolumn}
\usepackage{bm}
\usepackage[cmex10]{amsmath}

\begin{document}

\preprint{APS/123-QED}

\title{Magnetic field sensor with voltage-tunable sensing properties}

\author{Witold Skowro\'{n}ski}
 \email{skowron@agh.edu.pl}
\author{Piotr Wi\'{s}niowski}
\author{Tomasz Stobiecki}
\affiliation{Department of Electronics, AGH University of Science and Technology, Al. Mickiewicza 30, 30-059 Krak\'{o}w, Poland
}

\author{Sebastiaan van Dijken}
\affiliation{NanoSpin, Department of Applied Physics, Aalto University School of Science, P.O.Box 15100, FI-00076 Aalto, Finland}

\author{Susana Cardoso}
\author{Paulo P. Freitas}
\affiliation{INESC-MN and IN- Institute of Nanoscience and Nanotechnology, Lisbon 1000-029, Portugal}

\date{\today}

\begin {abstract}
We report on a magnetic field sensor based on CoFeB/MgO/CoFeB magnetic tunnel junctions. 
By taking advantage of the perpendicular magnetic anisotropy of the CoFeB/MgO interface, the magnetization of the sensing layer is tilted out-of-plane which results in a linear response to in-plane magnetic fields. The application of a bias voltage across the MgO tunnel barrier of the field sensor affects the magnetic anisotropy and thereby its sensing properties. An increase of the maximum sensitivity and simultaneous decrease of the magnetic field operating range by a factor of two is measured. Based on these results, we propose a voltage-tunable sensor design that allows for active control of the sensitivity and the operating filed range with the strength and polarity of the applied bias voltage.
\end{abstract}

\maketitle

Magnetic tunnel junctions (MTJs) based on CoFeB electrodes separated by a MgO tunneling barrier are excellent candidates for magnetic memory cells \cite{takemura_32-mb_2010}, magnetic field sensors and microwave nano-electronics components \cite{dussaux_large_2010, skowronski_zero-field_2012}, due to a high tunneling magnetoresistance (TMR) ratio and easy control of their magnetization with either magnetic field or spin polarized current. Magnetic field sensors utilizing TMR or giant magnetoresistance (GMR) effects are often designed with an orthogonal alignment between the sensing and reference layer magnetization \cite{tondra_picotesla_1998, liu_magnetic_2002, lu_shape-anisotropy-controlled_1997, lacour_field_2002}. This cross configuration produces a linear and reversible MR response by coherent rotation of the sensing layer magnetization in a perpendicular applied magnetic field. One viable way of realizing a magnetic cross geometry is the use of a sensing layer with out-of-plane magnetization \cite{van_dijken_magnetoresistance_2005}. The sensing properties of such sensors are usually fixed, and depending on the operating range or magnetic field sensitivity required for a certain application, different sensors needed to be utilized. 

Recently, it was shown that the magnetic anisotropy of thin ferromagnetic layers can be modified by an applied electric field \cite{weisheit_electric_2007, maruyama_large_2009, endo_electric-field_2010}. This feature was utilized in pseudo-spin valve structures to manipulate the MTJ resistance using bias voltages \cite{shiota_quantitative_2011} and it was used to demonstrate electric-field control of prototype memory cells \cite{shiota_induction_2011, wang_electric-field-assisted_2011}. In addition, it was shown that the resonance excitation of nanomagnets can be induced by electric fields alone \cite{nozaki_electric-field-induced_2012}.

In this letter, we demonstrate that electric-field control of magnetic anisotropy in cross-magnetization-geometry MTJs enables the design of magnetic field sensors with voltage-tunable sensing properties. Our experiments are conducted on exchanged-biased CoFeB/MgO/CoFeB MTJs with a synthetic antiferromagnet (SAF). In the sensors, the CoFeB sensing layer is oriented out-of-plane due to the perpendicular magnetic anisotropy of the MgO/CoFeB interface \cite{ikeda_perpendicular-anisotropy_2010, wisniowski_sensor_2012}. In combination with the in-plane magnetic anisotropy of the CoFeB reference layer (RL), this produces linear transfer curves. Moreover, both the operating field range and the field sensitivity are controllable by the strength and polarity of the applied bias voltage. 
The ability to actively modify the sensing characteristics without the need of extra signal lines or other materials opens the door to single adoptable field sensors that could replace the separate sensors that are typically required for highly sensitive detection and the measurements of large magnetic fields.

The thin-film magnetic field sensors of the following multilayer structure: Si /SiO$_2$ /Ta 5 /Ru 18 /Ta 3 /Pt$_{46}$Mn$_{54}$ 18 /Co$_{82}$Fe$_{18}$ 2.2 /Ru 0.9 /(Co$_{52}$Fe$_{48}$)$_{75}$B$_{25}$ 3 /MgO 1.35 /(Co$_{52}$Fe$_{48}$)$_{75}$B$_{25}$ 1.55 /Ru 5 /Ta 5 (thickness in nm) \cite{wisniowski_effect_2008} were deposited using Nordiko 2000 magnetron sputtering system. 
Before microfabrication processing, the films were passivated with a 15 nm of Ti$_{10}$W$_{90}$N$_{2}$ capping layer. The 1 $\times$ 1 inch wafers were patterned by direct write laser lithography and ion beam milling, which resulted in 540 devices with a tunnel junction dimension ranging from 1.5 $\times$ 3 $\mu$m up to 4 $\times$ 36 $\mu$m. The patterned wafer was annealed in high vacuum at 340$^\circ$C, for 1 hour in a magnetic field of 5 kOe. The MgO tunnel barrier thickness of 1.35 nm corresponded to a resistance area (RA) product of 90 k$\Omega\mu m^2$. 

Resistance vs. magnetic field loops are measured using a four-probe method. In this letter, positive voltage indicates electron transport from the bottom RL to the top sensing layer. The sensitivity of the sensor was measured directly using lock-in detection with synchronization to a weak (0.5 Oe) sinusoidal magnetic field on the top of the bias field.


\begin{figure}
	\begin{center}
		\includegraphics[width=8cm]{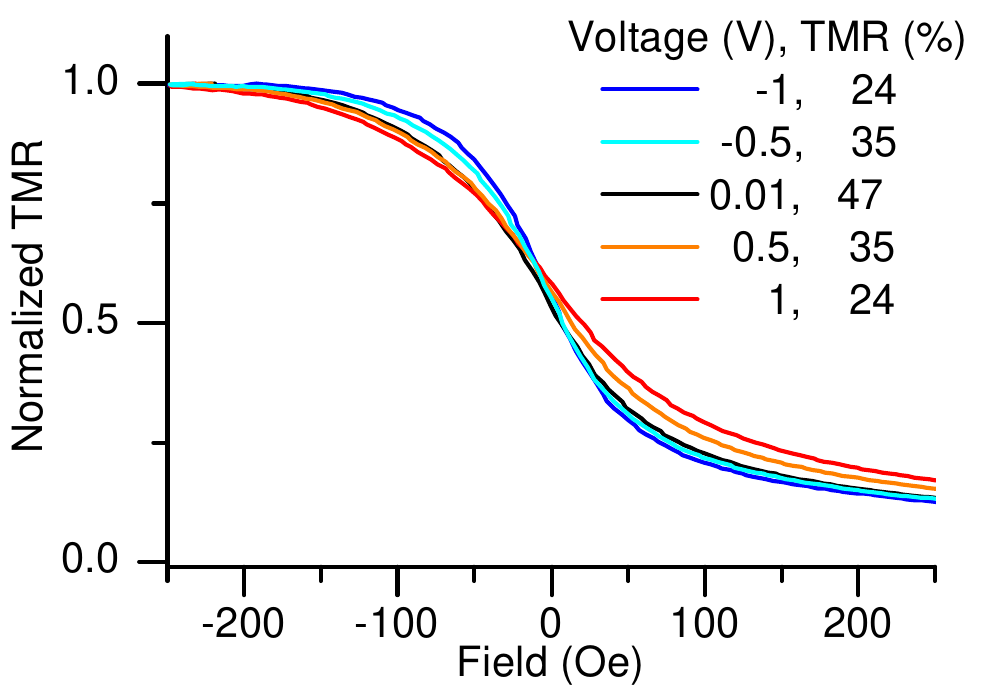}
	\end{center}
	\caption{Normalized TMR vs. in-plane magnetic field measured with different bias voltages. Changing the strength and polarity of the bias voltage alters the perpendicular magnetic anisotropy of the CoFeB sensing layer, which is reflected in the shape of the TMR curves.}
\label{fig:tmr_diff_vb}
\end{figure}

\begin{figure}
	\begin{center}
		\includegraphics[width=8cm]{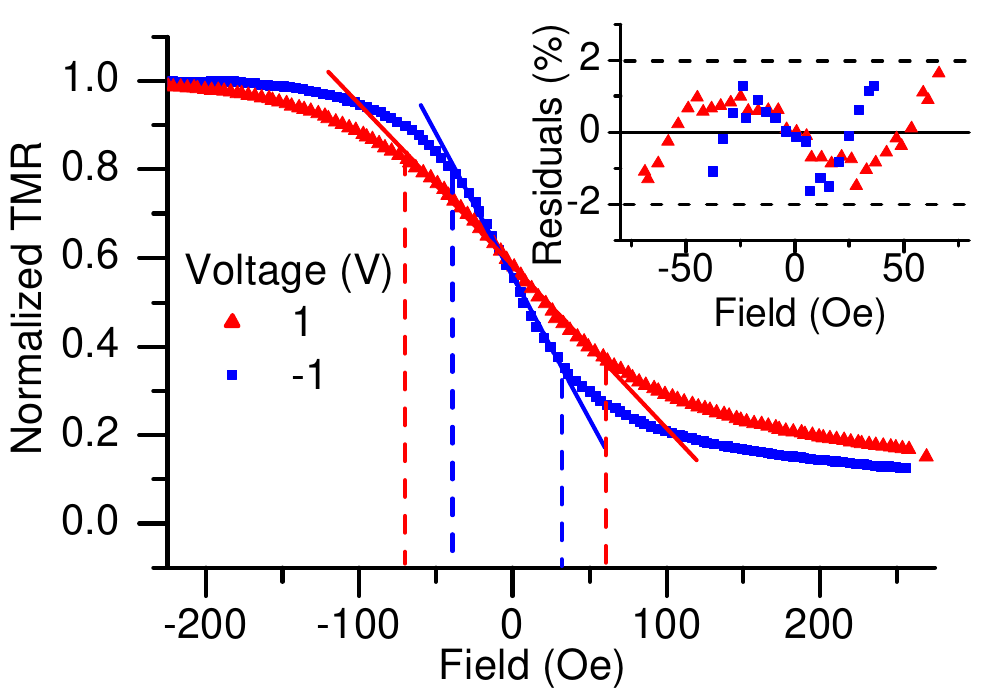}
	\end{center}
	\caption{Normalized TMR vs. in-plane magnetic field measured with $V_B$ = +/-1 V. Solid lines represent linear fits and dashed lines indicates linear operating regime. Within the operating field range, the field sensitivity calculated from the TMR curves is 0.082 \%/Oe and 0.149 \%/Oe for $V_B$ = 1 V and $V_B$ = -1 V, respectively. Inset: linear fit residuals of the sensor vs. magnetic field within the operating range.}
\label{fig:tmr_slope}
\end{figure}

Based on results presented in Fig. 1 of Ref. \cite{wisniowski_effect_2008}, we chose a CoFeB FL thickness of 1.55 nm, so that the resistance hysteresis loop is completely closed and linear in medium magnetic field ranges. This effect is caused by the perpendicular interface anisotropy of the thin CoFeB sensing layer on top of the MgO tunnel barrier \cite{ikeda_perpendicular-anisotropy_2010}. Figure \ref{fig:tmr_diff_vb} presents normalized TMR vs. in-plane magnetic field curves measured with different bias voltages ($V_B$) applied across the MgO tunnel barrier of a MTJ with an area of 4.5 $\mu$m$^2$. 

For the MTJ sensor, a linear and non-hysteretic resistance change is observed for all investigated $V_B$. The TMR ratio measured at low $V_B$ reaches 47\% and drops to 24\% for $V_B$ = 1 V. The polarity and magnitude of the applied bias voltage affects the magnetic anisotropy of the thin CoFeB sensing layer \cite{maruyama_large_2009, endo_electric-field_2010, shiota_induction_2011}. Positive bias voltages increase the out-of-plane magnetic anisotropy. As a consequence, the magnetization of the sensing layer rotates less when an in-plane magnetic field is applied and this increases the linear TMR response.   
Negative voltages reduce the out-of-plane anisotropy of the CoFeB layer and this results in a more sensitive sensor response. We note that the current density in the smallest MTJ sample only amounts 9 $\times$ 10$^2$ A/cm$^2$ under maximum bias conditions. This is three orders of magnitude smaller than the current density that is typically required for spin-transfer-torque (STT) induced resistance changes \cite{skowronski_interlayer_2010}. STT effects can therefore be excluded in this study.

\begin{figure}
	\begin{center}
		\includegraphics[width=8cm]{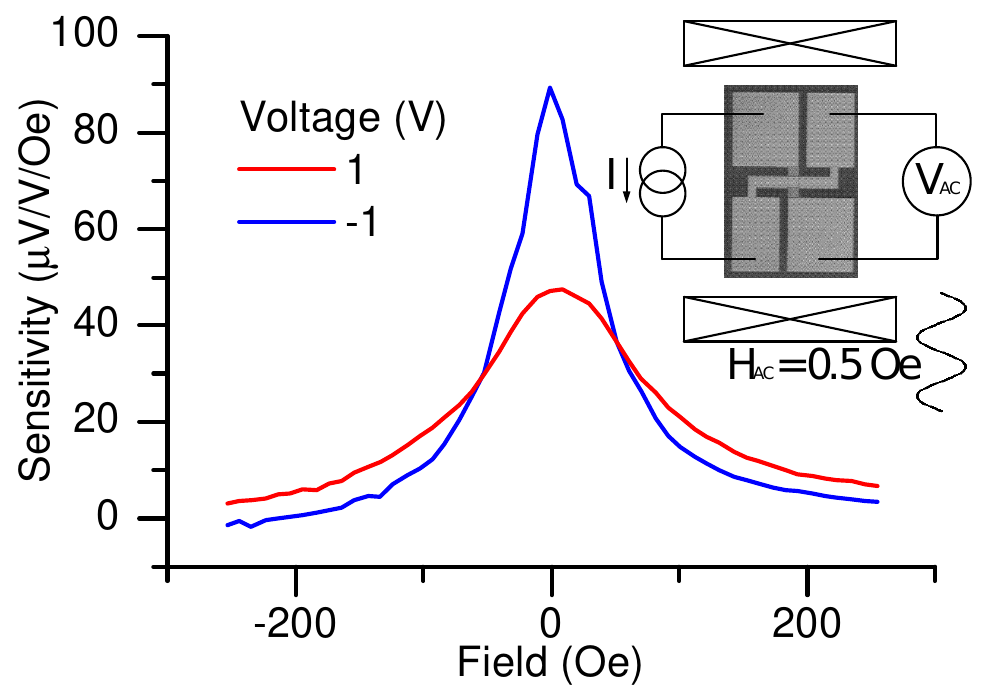}
	\end{center}
	\caption{Sensitivity of the sensor measured for $V_B$ = +/- 1V, using a four-probe and lock-in detection scheme (shown in the inset). The maximum sensitivity, measured for $H$ = 0 Oe, increases by a factor of two when the polarity of the 1 V bias voltage is reversed. }
\label{fig:sensitivity}
\end{figure}

Figure \ref{fig:tmr_slope} shows normalized TMR vs. in-plane magnetic field loops measured with $V_B$ = +/- 1V together with linear fits of the sensor response in small applied magnetic field. The linear operating range, defined as the field range in which the sensor response is linear within 2 \% error, changes from +/- 36 Oe for $V_B$ = -1 V, up to +/- 69 Oe for $V_B$ = 1 V. Similarly, the maximum field sensitivity, defined as the change of the TMR ratio divided by the linear operating range drops from 0.149 \%/Oe for $V_B$ = -1 V, to 0.082 \%/Oe for $V_B$ = 1 V. Similar results were obtained by direct sensitivity measurements as illustrated in Fig. \ref{fig:sensitivity}. In these experiments, the MTJ was placed in a sinusoidal magnetic field of $H_{AC}$ = 0.5 Oe. This produced an AC output signal, whose amplitude increases from 47 $\mu$V/V/Oe for $V_B$ = 1 V, up to 89 $\mu$V/V/Oe for $V_B$ = -1 V.
The MTJs with voltage-tunable TMR properties can be used as a smart sensors that operate in different field ranges. 
The breakdown voltage of the MTJs exceeds 2 V and, therefore, the changes in operating field range and sensitivity are obtained well below breakdown events. The other sensors on the wafer with different junction areas exhibit the same behavior, which is consistent with theoretical predictions that electric-field-induced changes are independent of the junction size.

\begin{figure}
	\begin{center}
		\includegraphics[width=8cm]{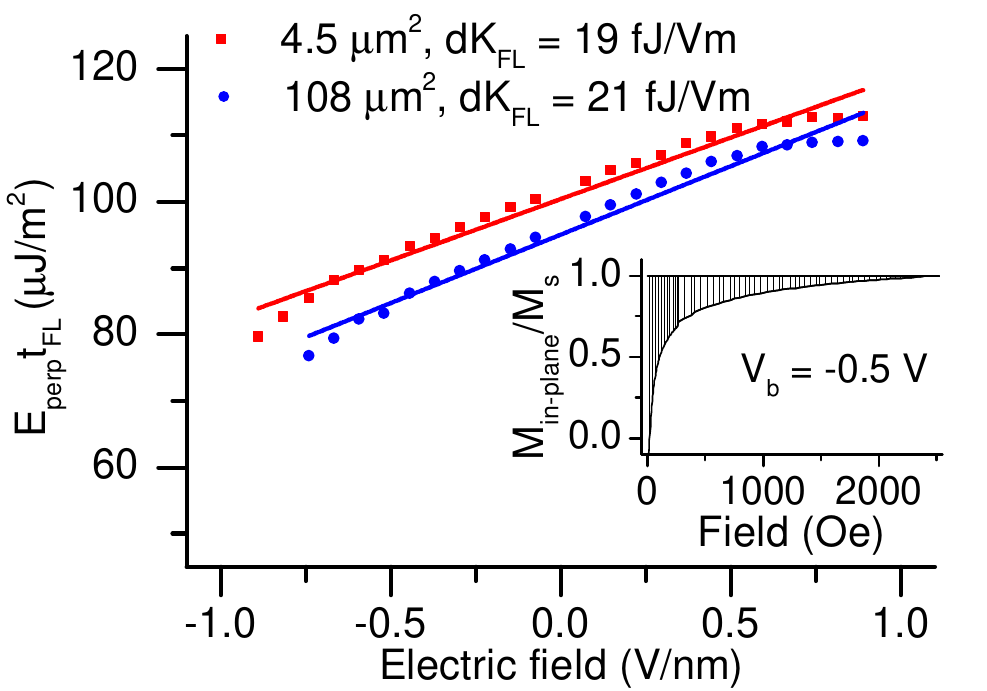}
	\end{center}
	\caption{Perpendicular magnetic anisotropy energy of the CoFeB sensing layer as a function of applied electric field. The data was extracted by integrating $M_{in}/M$ vs. magnetic field curves for MTJs with an area of 4.5 and 108 $\mu$m$^2$ (inset). }
\label{fig:anis_energy}
\end{figure}

In order to estimate the change of perpendicular magnetic anisotropy in an applied electric field, we performed an analysis that is similar to the one presented in Ref. \cite{shiota_quantitative_2011}. In this procedure, the ratio of the in-plane component of the magnetization $M_{in}$ to the total magnetization $M$ was estimated from the dependence of the resistance ($R$) on the angle $\theta$ between the magnetization of the sensing layer and RL:
\begin{equation}
	\label{eq:r_angle}
	R = R_p + \frac{R_{ap}-R_p}{2}\left(1-\frac{M_{in}}{M}\right)
\end{equation}
where, $R_p$ and $R_{ap}$ are the resistances for the parallel and antiparallel magnetization states, respectively, and $M_{in}$/$M$ = $\cos \theta$. The perpendicular anisotropy energy was calculated by integrating $M_{in}/M$ over the measured field range from 0 to 2500 Oe, which corresponds to the in-plane saturation field of the CoFeB sensing layer, for each $V_B$. An example of the integration area for $V_B$ = -0.5 V is presented in the inset of Fig. \ref{fig:anis_energy}. In our calculation, we used $\mu_0M_s$ = 0.93T for 1.55 nm thick (Co$_{52}$Fe$_{48}$)$_{75}$B$_{25}$ sensing layer.

As shown in Fig. \ref{fig:anis_energy} the perpendicular magnetic anisotropy energy of a 4.5 $\mu$m$^2$ sensor without $V_B$ applied is $E_{perp} t_{FL}$ = 101 $\mu$J/m$^2$. This energy changes approximately linearly with applied electric field. The linear field dependence corresponds to 19 fJ/Vm, which is in good agreement with sputter deposited and annealed CoFeB layers \cite{endo_electric-field_2010} and crystalline CoFe \cite{shiota_quantitative_2011}. For sensors with larger junction areas, similar values are obtained.

In summary, we have demonstrated a new magnetic field sensor concept with a voltage-tunable measurement range and field sensitivity. The sensor consists of a MgO-based magnetic tunnel junction with a thin CoFeB sensing layer exhibiting perpendicular magnetic anisotropy. The anisotropy strength depends linearly on the bias voltage across the MgO tunnel barrier. As a result, a wide range of sensing properties can be realized with one integrated device. In our proof of concept experiments, the measurement range and field sensitivity are enhanced by a factor of two when the polarity of the bias voltage is switched. This unique features are relevant for new magnetic field sensor designs with multiple integrated functions.

T.S. and W.S. acknowledge Foundation for Polish Science MPD Programme co-financed by the EU European Regional Development Fund. This work was supported by the Polish National Science Center grants (N505489040, 11.11.120.614 and Harmonia-2012/04/m/ST/00799), FCT through the PIDDAC Program and Portuguese National Project PTDC/CTM-NAN/110793/2009. S.V.D. acknowledges financial support from the Academy of Finland (SA 127731).

\bibliographystyle{nature}

\begin{thebibliography}{10}

\bibitem{takemura_32-mb_2010}
Takemura, R., Kawahara, T., Miura, K., Yamamoto, H., Hayakawa, J., Matsuzaki,
  N., Ono, K., Yamanouchi, M., Ito, K., Takahashi, H., Ikeda, S., Hasegawa, H.,
  Matsuoka, H., and Ohno, H.
\newblock {\em {IEEE} Journal of {Solid-State} Circuits}{ \bf 45}(4), 869--879
  (2010).

\bibitem{dussaux_large_2010}
Dussaux, A., Georges, B., Grollier, J., Cros, V., Khvalkovskiy, A., Fukushima,
  A., Konoto, M., Kubota, H., Yakushiji, K., Yuasa, S., Zvezdin, K., Ando, K.,
  and Fert, A.
\newblock {\em Nature Communications}{ \bf 1}(1), 1--6 (2010).

\bibitem{skowronski_zero-field_2012}
Skowro\'nski, W., Stobiecki, T., Wrona, J., Reiss, G., and van Dijken, S.
\newblock {\em Applied Physics Express}{ \bf 5}(6), 063005 May  (2012).

\bibitem{tondra_picotesla_1998}
Tondra, M., Daughton, J.~M., Wang, D., Beech, R.~S., Fink, A., and Taylor,
  J.~A.
\newblock {\em Journal of Applied Physics}{ \bf 83}(11), 6688--6690 (1998).

\bibitem{liu_magnetic_2002}
Liu, X., Ren, C., and Xiao, G.
\newblock {\em Journal of Applied Physics}{ \bf 92}(8), 4722--4725 (2002).

\bibitem{lu_shape-anisotropy-controlled_1997}
Lu, Y., Altman, R.~A., Marley, A., Rishton, S.~A., Trouilloud, P.~L., Xiao, G.,
  Gallagher, W.~J., and Parkin, S. S.~P.
\newblock {\em Applied Physics Letters}{ \bf 70}(19), 2610--2612 (1997).

\bibitem{lacour_field_2002}
Lacour, D., Jaffres, H., Dau, F. N.~V., Petroff, F., Vaures, A., and Humbert,
  J.
\newblock {\em Journal of Applied Physics}{ \bf 91}(7), 4655--4658 (2002).

\bibitem{van_dijken_magnetoresistance_2005}
van Dijken, S. and Coey, J. M.~D.
\newblock {\em Applied Physics Letters}{ \bf 87}(2), 022504 (2005).

\bibitem{weisheit_electric_2007}
Weisheit, M., Fahler, S., Marty, A., Souche, Y., Poinsignon, C., and Givord, D.
\newblock {\em Science}{ \bf 315}(5810), 349--351 January  (2007).

\bibitem{maruyama_large_2009}
Maruyama, T., Shiota, Y., Nozaki, T., Ohta, K., Toda, N., Mizuguchi, M.,
  Tulapurkar, A.~A., Shinjo, T., Shiraishi, M., Mizukami, S., Ando, Y., and
  Suzuki, Y.
\newblock {\em Nature Nanotechnology}{ \bf 4}(3), 158--161 January  (2009).

\bibitem{endo_electric-field_2010}
Endo, M., Kanai, S., Ikeda, S., Matsukura, F., and Ohno, H.
\newblock {\em Applied Physics Letters}{ \bf 96}(21), 212503 (2010).

\bibitem{shiota_quantitative_2011}
Shiota, Y., Murakami, S., Bonell, F., Nozaki, T., Shinjo, T., and Suzuki, Y.
\newblock {\em Applied Physics Express}{ \bf 4}(4), 043005 March  (2011).

\bibitem{shiota_induction_2011}
Shiota, Y., Nozaki, T., Bonell, F., Murakami, S., Shinjo, T., and Suzuki, Y.
\newblock {\em Nature Materials}{ \bf 11}(1), 39--43 November  (2011).

\bibitem{wang_electric-field-assisted_2011}
Wang, W.-G., Li, M., Hageman, S., and Chien, C.~L.
\newblock {\em Nature Materials}{ \bf 11}(1), 64--68 November  (2011).

\bibitem{nozaki_electric-field-induced_2012}
Nozaki, T., Shiota, Y., Miwa, S., Murakami, S., Bonell, F., Ishibashi, S.,
  Kubota, H., Yakushiji, K., Saruya, T., Fukushima, A., Yuasa, S., Shinjo, T.,
  and Suzuki, Y.
\newblock {\em Nature Physics}{ \bf 8}(6), 492--497 April  (2012).

\bibitem{ikeda_perpendicular-anisotropy_2010}
Ikeda, S., Miura, K., Yamamoto, H., Mizunuma, K., Gan, H.~D., Endo, M., Kanai,
  S., Hayakawa, J., Matsukura, F., and Ohno, H.
\newblock {\em Nature Materials}{ \bf 9}(9), 721--724 (2010).

\bibitem{wisniowski_sensor_2012}
Wi\'{s}niowski, P., Wrona, J., Stobiecki, T., Cardoso, S., and Freitas, P.~P.
\newblock {\em {IEEE} Transactions on Magnetics}{ \bf 48}(11), 2450 (2012).

\bibitem{wisniowski_effect_2008}
Wi\'{s}niowski, P., Almeida, J.~M., Cardoso, S., Barradas, N.~P., and Freitas,
  P.~P.
\newblock {\em Journal of Applied Physics}{ \bf 103}(7), 07A910 (2008).

\bibitem{skowronski_interlayer_2010}
Skowro\'{n}ski, W., Stobiecki, T., Wrona, J., Rott, K., Thomas, A., Reiss, G.,
  and van Dijken, S.
\newblock {\em Journal of Applied Physics}{ \bf 107}(9), 093917 (2010).

\end{thebibliography}

\end{document}